\documentclass[12pt,preprint]{aastex}

\def\ltsima{$\; \buildrel < \over \sim \;$}
\def\gtsima{$\; \buildrel > \over \sim \;$}
\def\lsim{\lower.5ex\hbox{\ltsima}}
\def\gsim{\lower.5ex\hbox{\gtsima}}
\def\lapp{\ifmmode\stackrel{<}{_{\sim}}\else$\stackrel{<}{_{\sim}}$\fi}
\def\gapp{\ifmmode\stackrel{>}{_{\sim}}\else$\stackrel{<}{_{\sim}}$\fi}

\newdimen\minuswidth    
\setbox0=\hbox{$-$}
\minuswidth=\wd0
\catcode`@=\active
\def@{\kern\minuswidth}
\setbox0=\hbox{\rm0}
 
\shorttitle{The BSS population of the globular cluster M10.} 
\shortauthors{Dalessandro et al.}
 
\begin{document} 
\title{UV observations of the globular cluster M10 from HST and GALEX. The BSS population.\footnote{Based on observations
    collected with the NASA/ESA {\it HST}, obtained at the Space
    Telescope Science Institute, which is operated by AURA, Inc.,
    under NASA contract NAS5-26555. Also based on WFI observations collected at the 
    European Southern Observatory, La Silla, Chile,
    within the observing program 69.D-0582.}  }

\author{
E. Dalessandro\altaffilmark{2},
F. R. Ferraro\altaffilmark{2}, 
B. Lanzoni\altaffilmark{2},
R. P. Schiavon\altaffilmark{3},
R. W. O'Connell\altaffilmark{4}
and G. Beccari\altaffilmark{5}
}
\affil{\altaffilmark{2} Dipartimento di Astronomia, Universit\`a degli Studi
di Bologna, via Ranzani 1, I--40127 Bologna, Italy}
\affil{\altaffilmark{3} Astrophysics Research Institute, Liverpool John Moores University,
 Twelve Quays House, Egerton Wharf, Birkenhead CH41 1LD, UK}
\affil{\altaffilmark{4} Astronomy Department, University of Virginia, P.O. Box 400325,
	Charlottesville, VA 22904, USA}
\affil{\altaffilmark{5} ESO - European Southern Observatory,
  Karl-Swarzschild Str. 2, D-85748 Garching bei M\"unchen,  Germany} 
\date{03 May, 2013}

\begin{abstract}
We present a combination of high-resolution {\it Hubble Space
Telescope} and wide-field ground-based and {\it Galaxy Evolution
Explorer} data of the Galactic Globular Cluster M10 (NGC6254).  By
using this large data-set we determined the center of gravity of
the cluster and we built its density profile from star counts
over its entire radial extension. We find that the density profile
is well reproduced by a single-mass King model with structural
parameters $c=1.41$ and $r_c=41\arcsec$.  We also studied the Blue
Straggler Star population and its radial distribution.  We count a
total number of 120 BSS within the tidal radius.  Their radial
distribution is bimodal: highly peaked in the cluster
center, decreasing at intermediate distances and rising again outwards.
We discuss these results in the context of the {\it dynamical clock}
scheme presented by Ferraro et al. (2012) and of recent results
about the radial distribution of binary systems in this cluster.
\end{abstract}


\section{INTRODUCTION}

Blue Stragglers Stars (BSS) are a common population of any medium-size
stellar aggregate such as open (Mathieu \& Geller 2009) and globular clusters, as well
as dwarf galaxies (Monelli et al. 2012). They are arguably the most common objects
whose nature is not explainable in terms of the canonical evolution
of a single star.\\ In an optical Color Magnitude Diagram (CMD) BSS
appear hotter and brighter than Turn-Off (TO) stars, thus mimicking
a sparse sequence of younger and more massive objects. Observational
evidence (see for example Shara et al. 1997; Gilliland et al.
1998) showed that BSS are indeed more massive ($M \sim 1.2 - 2.0
M_{\odot}$) than ``normal'' stars from the same cluster ($M
\sim 0.8 M_{\odot}$) thus they are thought to be the result of some
mechanism responsible for increasing the masses of single
stars. \\ Two main formation scenarios have been proposed over the
years: the "collisional" scenario (COL-BSS; Hills \& Day 1976)
according to which BSS are the end-products of stellar mergers
induced by collisions  between single stars or binary systems;
and the "mass-transfer scenario" (MT-BSS; McCrea 1964; Zinn \&
Searle 1976) in which BSS are the result of mass accretion between
two stars in a binary system.  These mechanisms are believed to
work simultaneously within the same cluster (see the case of M30;
Ferraro et al. 2009), with efficiencies that may be a function
of environment (Ferraro et al. 1995; Davies et al. 2004). In particular, COL-BSS have higher probabilities
to form in the cores of GCs where densities are incredibly high and
so is the chance of collisions and interactions. On the contrary,
the MT scenario is the dominant formation channel of BSS in very
loose systems or in cluster outskirts, where binary systems
are more likely to have evolved free of interactions
with other stars.  \\ Observational support to this idea
comes from the results obtained from data for 47~Tucanae by
Ferraro et al. (2006a; see also Lovisi et al. 2010 for M4), where
a fraction of BSS has been found to have strong Carbon (C) and
Oxygen (O) abundance anomalies probably due to mass transfer processes
(Sarna \& de Greve 1996). \\ Being 3-4 times more massive (independently of the formation mechanism) 
than the average cluster mass ($m\sim0.3M_{\odot}$), BSS are heavily affected by dynamical friction and thus
they are natural test particles to probe the internal dynamics of stellar aggregates. In particular
their radial distribution (Ferraro et al.  1997) has been
found to be an important tool to probe the efficiency of dynamical
friction (Mapelli et al. 2006; Lanzoni et al.  2007a).  This approach
has been followed by our group over the past fifteen years
leading to the construction of a large collection of BSS
radial distributions (see Dalessandro et al. 2008a and references
therein) over the full spatial extent of the clusters studied.
\\ By using this data set, Ferraro et al.  (2012; hereafter
F12) introduced the definition of the {\it "dynamical clock"}, i.e.
a measurement that provides an assessment of the dynamical
age of a cluster on the basis of the presence and the position
of the minimum of its BSS radial distribution.  \\ Following
this interpretation, clusters with a flat radial distribution
($\omega$Centauri - Ferraro et al.  2006b; NGC2419 - Dalessandro
et al. 2008b; Palomar14 - Beccari et al. 2011) are dynamically
young, so that the effect of dynamical friction is still not
present. Clusters whose BSS distribution has a
single peak at the center (M79 - Lanzoni et al.  2007b; M75 -
Contreras Ramos et al. 2012; M30 - Ferraro et al. 2009; M80 - F12) are
dynamically old, so that the most remote BSS already
have drifted towards the cluster center.  Finally intermediate
dynamical-age clusters show a bimodal BSS radial distribution with
a peak in the cluster center and another one towards the outer
regions, thus defining a clear minimum in the distribution. The
radial position of this minimum ($r_{min}$) has been found to be
linked to the the efficiency of the dynamical friction and to
strongly correlate with the central relaxation time (F12).\\ In
this work we present a photometric analysis of the GGC M10 (NGC
6254) with the aim of studying its structural parameters and the
properties of its BSS population. The results will be 
interpreted in the framework of the scheme proposed by F12
and will be compared with previous results on M10's binary
systems content and mass segregation profile (Beccari et al.
2010, hereafter B10; Dalessandro et al.  2011, hereafter D11).

\section{DATA ANALYSIS}

As suggested by Ferraro et al. (1997; 2004), since BSS are relatively
warm stars ($T_{eff}\sim7000-10000 K$), the optimal way to observe
them is at ultraviolet (UV) wavelengths, where they  define a
bright and well defined sequence that is easily distinguishable in
the CMD. In fact in the wavelength regime
between $1800\AA$ and $3500\AA$, giants, that are the brightest
objects at optical and near infrared (IR) bands, give a negligible
contribution to the total light.
Therefore, stellar crowding is negligible, which makes identification
of hot stars very easy.

\subsection{Observations and data reduction}

We combined high resolution optical and UV images obtained with HST
and optical and UV wide field images collected both from ground-based
and space telescopes.  In particular the data-set is composed of:
\begin{enumerate} \item{{\it high-resolution sample} --  It consists
of a combination of images obtained with the Advanced Camera for
Surveys (ACS) and the Wide Field Camera 2 (WFPC2) on board the
Hubble Space Telescope (HST).  The WFPC2 data set (Prop. 6607, PI:
F.R.Ferraro) has been already presented by Ferraro et al. (2003)
and it consists of a series of images in the F255W, F336W and F555W
filters.
\\
This data-set has been supplemented by ACS optical images
already used by B10 and D11. In addition, in order to minimize the
impact of saturation on bright stars, we supplemented that
catalogue with short exposure time images: two images in F606W and
F814W with exposure time $t_{exp}=7 sec$ each (Prop: 10775, PI:
Sarajedini).\\ As shown in Figure~1, the centre of the cluster is
located in the Planetary Camera (PC) for the WFPC2 data-set and in
the gap between the two chips of the ACS data-set.\\

\item{{\it wide field sample} -- It consists of UV wide-field images secured with 
the {\it Galaxy Evolution Explorer} (GALEX;
Schiavon et al. 2012; Dalessandro et al. 2012) and optical ground-based wide-field images obtained
 with the Wide Field Imager (WFI) mounted at the 2.2m ESO-MPG telescope.\\
 The GALEX data set (PI R.P. Schiavon, GI1 release) has been already presented by Schiavon et al. (2012)
 and it consists 
 of two circular images of $\sim 1\deg^2$, one obtained with the FUV channel and $t_{exp}=1911 sec$ and
 one with NUV channel with $t_{exp}=25362 sec$.
 These images are aligned and the core of the cluster is approximately located in the centre of the detectors. \\
 The WFI data-set (Prop. 69.D-0582(A), PI Ortolani) is composed of 9 images in the I band, 6 with $t_{exp}=30 sec$
 and  3 with $t_{exp}=3 sec$ and of 9 V band images, 6 with $t_{exp}=50 sec$ and 3 with $t_{exp}=5 sec$.}}

\end{enumerate}

The photometric analysis has been carried out by using 
DAOPHOTII (Stetson et al. 1987) for all the images with the exception of the WFPC2 data-set 
for which ROMAFOT (Buonanno et al. 1983) has been adopted. For the full description of the photometric 
reduction procedure see
Ferraro et al. (2003), B10 and Schiavon et al. (2012). 

\subsection{Astrometry and photometric calibration.}

The catalogues obtained from each data-set were put on the absolute astrometric system
using a large number of stars in common with the {\it Guide Star Catalogue} (GSC2.3) .
First, we obtained the astrometric solution for each of the 8 WFI chips
by using a third order polynomial solution found by the 
{\rm CataXcorr} package\footnote{CataXcorr is a
catalog cross-correlation software developed at the Bologna Observatory by P. Montegriffo}.
Since the core regions of GC do not host astrometric standard stars, we used the stars in the WFI catalogue 
as secondary astrometric standards to extend the astrometric solution to the internal regions.
All the stars in common with GALEX, and with the {\it high resolution sample} were then used as secondary
astrometric standards.
Several hundreds of them have been found in each step and this allowed 
us to find a very accurate astrometric solution with typical uncertainties of about $0.2\arcsec$ in both 
right ascension ($\alpha$) and declination ($\delta$).\\

More than 200 stars in common with the publicly available 
"Photometric Standard Fields"\footnote{http://www3.cadc-ccda.hia-iha.nrc-cnrc.gc.ca/community/STETSON/standards/}
catalogue (Stetson 2000) were used to calculate the color equations needed to transform the
V and I WFI magnitude into the standard Johnson-Cousin photometric system.
The ACS $m_{F606W}$ and $m_{F814W}$ magnitudes were
calibrated respectively to V and I by using the stars in common
with the WFI catalog. In particular, the best linear fits to the
star distribution in the (V-$m_{F606W}$, V-I) or the (I-$m_{F814W}$, V-I) planes
have been applied.  The same approach has been adopted for the
$m_{F555W}$ WFPC2 magnitudes, while $m_{F255W}$ and $m_{F336W}$
have been calibrated to the STMAG photometric system by using the
procedure and Zero-Points reported in Holtzman et al. (1995). 
The NUV ABMAG GALEX magnitudes have been converted to $m_{F255W}$ by
means of Equation~1 obtained by linearly fitting the
stars in common in the ($m_{F255W}-NUV, m_{F255W}-V$) plane. 
\begin{equation}
m_{F255W}= NUV + 0.02\times(m_{F255W}-V) + 0.39
\end{equation}
The final catalogue is composed of all stars in the WFPC2
catalogue and those detected in the complementary ACS FOV (see
Figure 1).  It also includes the stars in the WFI catalogue and
external to both the WFPC2 and ACS FOVs (Figure~2).  Stars
in the WFI catalogue and in common with GALEX have also a $m_{F255W}$
magnitude.

\subsection{Center of gravity and density profile.}

We determined the center of gravity $C_{grav}$ of M10 by averaging
the positions $\alpha$ and $\delta$ of selected stars (see below)
lying in the FOV of the WFPC2/Planetary Camera, as
previously done in other works by our group (see Dalessandro et al.
2009 for example).  We choose the center listed by Goldsbury et al. (2010)
as  initial guess for our iterative procedure (Montegriffo et
al. 1995).  In order to avoid spurious or incompleteness effects,
we performed three different  measurements by using three sub-sample of
stars respectively with $V<19$, $V<18.5$ and $V<18$. The resulting
$C_{grav}$ is the average of three  measurements thus obtained and it
is located at $\alpha_{J2000}=16^h57^m8^s.92$,
$\delta_{J2000}=-4^{\circ} 05\arcmin 58\arcsec.07$ ($RA=254.2871636$
$Dec=-4.0994655$). $C_{grav}$ is located at $\sim3.5\arcsec$ North-West from
the one published by Goldsbury et al. (2010). We checked that this difference
does not have any impact on the present analysis.\\

Taking advantage of the wide radial coverage of our photometric
catalogue, which extends to distances from $C_{grav}$  to $r\sim
1400\arcsec$, we constructed the star count density
profile for the full cluster extent. We considered only
stars with $14<V<19$ in the different data-sets (see Figure~3).  We
divided the total FOV in 28 concentric annuli centered on $C_{grav}$
and reaching $r=1300\arcsec$.  Each annulus was then split in a
number of sub-sectors (2 or 4 depending on the maximum angular coverage constrained by the WFI FOV).  
In each sub-sector the density  was estimated as the
ratio between star counts and area. The
density assigned to a given annulus is the average of the densities
computed in each sub-sector.  The error assigned to each density
measure is defined as the dispersion from the mean of the sub-sectors
composing the considered annulus.  The resulting density profile
is shown in Figure.~4. The three outermost points have almost the
same density, defining a plateau, which was used to
define the background density level ($\sim 2
stars/arcmin^2$).\\ We performed a fit of the profile by using a
single-mass King model (King 1966). The best fit model is shown in
Figure~4 and the derived values of the core radius $r_c=41\pm1\arcsec$,
concentration $c=1.41\pm0.03$ and tidal radius
$r_t=1053\pm100\arcsec$\footnote{In the following analysis we will
adopt as $r_t$ the upper limit to the value obtained by the best
fit King model.} are reported.  The concentration obtained from our best-fit 
is in good agreement with those found by Harris (1996 - edition 2010; $c=1.40$)
and McLauglhin \& Van der Marel (2005; $c=1.41$), while $r_c$
is slightly smaller than previous estimates ($r_c=51\arcsec$ from Harris 1996;
$r_c=46.4\arcsec$ from McLauglhin \& Van der Marel 2005) based on surface 
brightness profiles fitting.
In an upcoming paper (Miocchi et al. 2013, in preparation), we will present a detailed comparison between observed density
profiles and different theoretical models for a large sample of GGCs. For the case of M10,
multi-mass King models and Wilson models (Wilson 1975)
result in a significantly lower quality fit than the best solution described above.

\section{SAMPLE DEFINITION}
In order to study the BSS radial distribution, we need to select both 
BSS and at least one reference star population. 

\subsection{BSS selection}

As explained in Section~2, CMDs involving UV photometry provide
the best means for one to study the hottest stellar populations
like BSS in GCs. In particular, the $m_{F255W}$ filter from
HST/WFPC2 allows for an excellent contrast between BSS and main
sequence turnoff stars (see also Figure 1 in Ferraro 2006c).
In order to make the selection between different samples as homogeneous
as possible, we used the ($m_{F255W},m_{F255W}-V$) CMD. In this way, 
selection of BSS from WFPC2 and WFI/GALEX
photometry was based on the same criteria. \\ As shown in Figures~5 and 6,
in order to avoid possible contamination from TO and Sub Giant
Branch (SGB) stars, we adopted a limiting magnitude $m_{F255W}=20$
(that is $\sim 1$ mag brighter than the cluster TO). In addition,
in order to limit the contamination from possible blends with the
Red Giant Branch (RGB) stars, we restricted our sample to stars
bluer than ($m_{F255W}-V)=2.5$. The boxes shown in Figures~5 and 6
illustrate this selection.  These limits have been fixed on the
basis of the quality of both the WFPC2 and the WFI/GALEX NUV CMDs.
With these criteria we selected 47 and 48 BSS in the WFPC2 and
WFI/GALEX FOVs respectively.  \\ The complementary ACS is the only
catalogue for which we do not have UV band measurements.  We guarantee
the homogeneity of the BSS selection by translating the $m_{F255W}$
magnitude cut into the V band.  The adopted magnitude limit
$m_{F255W}=20$ corresponds to $V\sim18$.  We adopted this magnitude
cut to select BSS in the ACS FOV. Also in this case, an additional
constraint on the (V-I) color has been adopted ($(V-I)<0.8$). In
this way we selected 25 additional BSS in the complementary ACS FOV
(Figure~7).\\
We obtained a rough estimate of the possible Galaxy field star
contamination by counting the number of stars located in our catalogue
at distances larger than $r_t$ and lying in the adopted (V,V-I)
selection box.  We find only one object over an area of about
$150 arcmin^2$ (corresponding to $\sim0.01 stars/arcmin^2$).

\subsection{The Reference populations}
The selection of the RGB stars, adopted as first reference population,
has been done in the optical CMDs where they are the brightest
stars.  In particular we selected
all the stars with $13.8<V<17$ along the RGB mean ridge line (see Figure~7).
With these limits we selected
respectively 229, 203 and 538 RGBs in the WFPC2, ACS and WFI samples
(open pentagons in Figure~7).\\ As it is possible to see in Figure~7,
field stars describe a vertical sequence at $(V-I)\sim1$ thus
contaminating mainly the SGB and RGB sequences. As done for BSS we
counted the number of field stars in the RGB region at $r>r_t$.  We
found 36 objects, resulting in an average density of $\sim0.2
stars/arcmin^2$.  We took this fraction into account for the estimate
of the populations ratios (Section~4).\\

We used as second reference population the Horizontal Branch (HB)
stars. When possible, the selection was performed in the
($m_{F255W},m_{F255W}-V$) CMD as shown in Figures 5 and 6. We found in our
sample 90 HBs in the WFPC2 FOV, 86 in ACS and 129 HB in GALEX/WFI.
We count 3 field stars lying in the CMD in the HB selection area
and at $r>r_t$ corresponding to an average density of $\sim 0.02
stars/arcmin^2$.\\ In addition to the statistical decontamination
of the selected populations, we looked for extra-galactic sources
by matching our catalogue with the NASA Extragalactic Database
(NED)\footnote{http://ned.ipac.caltech.edu/}.  A total of 14 sources
has been found: 12 galaxies and 2 Quasars (QSOs) (see Figure~7 right
panel).  The Quasar [HGP92] 165429.30-040340.3 (source \# 2 in the
NED catalogue) is in the HB region (see also Figure.~6) and it has
been removed from the HB population.\\ We identified three
out of four known variable stars in M10 (Sawyer 1938; Arp 1955 and
Voroshilov 1971): V1, V2 are respectively a foreground WUma and a
background RRlyrae (von Braun et al. 2002), while V4 is an HB star
in the instability strip.  We could not match V3 with our catalogue,
since it lies in one of the WFI gaps (Figure~2).

\section{THE BSS RADIAL DISTRIBUTION.} 
In order to analyze the BSS
radial distribution we first compared its cumulative radial
distribution to those of the reference populations.\\ 
We took into account the effect of contamination by statistically decontaminating the selected populations.
In particular we divided
the FOV in 5 concentric annuli centered on $C_{grav}$ and for each
of them we randomly subtracted a number of stars constrained by the
average field densities quoted in the previous Sections. The
cumulative radial distribution of the statistically decontaminated
populations is shown in Figure~8.  As apparent, BSS (solid line)
are more centrally concentrated than RGBs and HBs (dashed and dotted
lines respectively).  We assessed the statistical significance of
the observed difference between the radial distribution of the three
populations by means of a Kolmogorov-Smirnov test. We found that
BSS and HB have a probability P=92\% (corresponding to $\sim1.7\sigma$)
to be extracted from different parent populations, and BSS and RGB
have $P>99.9\%$ (corresponding to a significance larger than
$3\sigma$).\\

For a more quantitative analysis, we divided the total FOV in the
same five annuli used before. The number of stars of each decontaminated 
population are reported in Table~1. In each of them we estimated the ratio
between the number of decontaminated BSS and reference population
stars in each annulus ($N_{BSS}^{ann}$/$N_{ref}^{ann}$).  The radial
distribution is clearly bimodal (see Figure 9), strongly peaked in
the central regions, decreasing at intermediate radii and rising
again outwards. As expected for "normal" cluster populations, the
number ratio between the reference populations $N_{HB}/N_{RGB}$
does not show any peculiar and significant trend.\footnote{HB stars
are mildly more centrally concentrated than RGBs for $r<100\arcsec$
(Figure~9 and see also Figure~10).  This is also observable looking
at their relative cumulative radial distributions (Figure~8). However
the statistical significance of this difference is relatively small,
as highlighted earlier in this Section.} As discussed in the
Introduction, the BSS bimodal distribution is a common feature of
most of the GCs studied so far (see F12).\\ As a further check we
compared the distribution of the double normalized ratio (Ferraro
et al. 1993):

\begin{equation} 
R_{POP}=\frac{N_{POP}^{ann}}{N_{POP}^{tot}} /\frac{L_{samp}^{ann}}{L_{samp}^{tot}} 
\end{equation}

where $L_{samp}^{ann}$ and $L_{samp}^{tot}$ are the light sampled
in each annulus and in the field covered by the observations, respectively (Table~1).  \\
Figure~10 shows that the
distributions of HB and RGB stars follow that of the cluster
sampled light and $R_{HB/RGB}=1$, as expected by stellar  evolution
theory (Renzini \& Buzzoni 1989) for post-MS stars, while the
distribution of $R_{BSS}$  is bimodal.  The normalized ratio
at the cluster center is found to be $R_{BSS}\sim1.8$, which is in good
agreement with values found for the central peaks of other
GGCs (F12).  The minimum of the distribution is
located at $r_{min}\sim10.35r_c$.  In the cluster outskirts the
normalized ratio is
$R_{BSS}\sim1$, which is what one should expect to find for binary
systems and their by-products evolved in isolation (see Section 5)

\section{DISCUSSION.}

Simple analytical models (Mapelli et al. 2004, 2006) have
shown that the observed radial distribution of BSS is well reproduced
by modeling the dynamical friction acting on the binary population.
By using equation (1) in Mapelli et al. (2004; see also Binney \&
Tremaine 1987) and the best-fit King model obtained in Section~2.3,
we estimated the distance at which a star as massive as $1.2
M_{\odot}$ (estimated average mass for a BSS; Shara et al. 1997)
would have already sunk into the cluster core because of
dynamical friction. It is predicted to be $r\sim 11r_c$, nicely matching the observed
minimum in the BSS radial distribution.\\ 
As clearly stated in F12,
different BSS radial distributions are the observational evidence
of different stages of the cluster dynamical history. As time
passes, the distance from the cluster center at which dynamical
friction has been effective on BSS (and their progenitors) increases,
and so does the position of the minimum in the observed radial
distribution.  In particular, when mass segregation and dynamical
friction start to segregate binaries (and their by-products) into
the cluster centre, a central peak in the BSS distribution occurs.
Since the action of dynamical friction and mass segregation
progressively extends to larger and larger distances from the centre,
the dip left by the massive objects sunk to the bottom of the
potential becomes visible at progressively increasing radii. In the
meanwhile, the most remote BSS are still evolving in isolation in
the outer regions of the cluster with nearly the same initial
frequency (this generates the rising branch of the observed bimodal
BSS distributions).  Hence because of the dynamical friction effect,
the dip of the distribution progressively moves outward with
increasing (dynamical) age of the cluster.  In a highly evolved
cluster we can expect that virtually all the binaries and their by-products (at any
distance from the cluster centre) have sunk into the cluster core,
thus generating a single central peak in the BSS radial distribution,
as observed for example in NGC1904 (Lanzoni et al. 2007b).\\ F12
show that the BSS radial distribution is a good tool to rank the
dynamical stage reached by stellar systems, thus allowing a direct
measure of cluster dynamical age purely from observed properties.
In F12, M10 is classified as a {\it Family II} GC. This family is
populated by clusters with intermediate dynamical ages.  In addition,
the $r_{min}$ value found for M10 positions it towards the
dynamically older end of {\it Family II} clusters.  F12 showed
that the core relaxation time (as well as the half-mass relaxation
time) is a linear function of $r_{min}$. This behavior allowed the
definition of the {\it "dynamical clock"}: clusters with relaxation times
of the order of the age of the Universe show no sign of BSS segregation
(hence the radial distribution is flat and $r_{min}$ is not defined).
For decreasing relaxation times the radial position of
the minimum increases progressively. \\
The structural parameters obtained in the present work for M10 are
slightly different from those adopted by F12.  Following the approach
used by the authors, we detail here the calculations for the core
relaxation time and the position of M10 in the {\it "dynamical clock"}
scheme on the basis of the results obtained in Section~2.3.\\ The
core relaxation time ($t_{\rm rc}$) was computed by using
equation (10) in Djorgovski (1993): 
\begin{equation} 
t_{rc} = 1.491 \times 10^7 {\rm yr} \times \frac{0.5592}{\ln(0.4 N_\star)}
\frac{\sqrt{\rho_{M,0}}}{\langle m_\star\rangle} r_c^3, 
\end{equation}
where $N_\star$ is the total number of stars in the cluster,
$\rho_{M,0}$ is the central mass density in $M_\odot/$pc$^3$ and
$r_c$ is the core radius in pc.  The values quoted in Djorgovski
(1993) for a number of parameters were adopted, namely: the
average stellar mass $\langle m_\star\rangle= 0.3 M_\odot$, the
stellar mass-to-light ratio in the $V$-band $M/L_V=3$, the Sun
$V$-band absolute magnitude $M_{\odot V}=4.79$.  The number of stars
$N_\star$ has been obtained as the ratio between the total cluster
mass $M_{\rm tot}$ and $\langle m_\star\rangle$. 
$M_{\rm tot}$ has been calculated by using the $M/L_V$ value 
quoted above, a
bolometric correction factor for the $V$-band of 1.4 and the total luminosity
$L_V$. 
The latter has been derived from the integrated $V$
magnitude of the cluster $V_t=6.60$ (Harris 1996), and the distance
modulus $(M-m)_0=13.38$ and reddening $E(B-V)=0.28$ quoted in Ferraro
et al. (1999). The central mass density has been obtained as
$\rho_{M,0} = \mu_V/(p\times r_c)$, where $\mu_V$ is the central
surface brightness in $L_\odot$/pc$^2$, $p$ is a parameter which
depends on the cluster King concentration $c$ as in eq. (6) of
Djorgovski (1993). To compute these parameters we used the $\mu_V=17.70$
mag/arcsec$^2$ quoted by Harris (1996) and corrected for extinction,
and the concentration ($c=1.41$) and core radius ($r_c=41\arcsec$)
obtained from the best fit King model.


In Figure~11, we highlight the new position of M10 
in the ($t_{rc}/t_H$,$r_{min}$) logarithmic plane. As already discussed, M10 represents 
together with a few other GCs, a sort of bridge between {\it Family II} and {\it Family III} (triangles 
Figures~12). In fact $r_{min}$ is quite distant from the cluster center, but the BSS radial distribution
is still bimodal. \\
In recent studies about both the mass segregation profile (B10) and the radial 
distribution of binary stars (D11), we have obtained similar results 
to the ones reported in the present work.
In particular the radial distributions of binary stars and BSS show a very similar behavior at 
least at distances in which binaries have been analyzed ($r<300\arcsec$).
Indeed they both show clear evidence that in M10 energy equipartition 
has been achieved and heavier stars are moving on inner orbits. 

\acknowledgments 
We thank the anonymous referee for the careful reading and useful comments that improved 
the presentation of this work.
This research is part of the project {\it COSMIC-LAB} (http://www.comic-lab.eu)
funded by the {\it European Research Council} (under contract
ERC-2010-AdG-267675). RPS acknowledges support from the Gemini Observatory, which is operated by the Association 
of Universities for Research in Astronomy, Inc., on behalf of the international Gemini partnership of Argentina, 
Australia, Brazil, Canada, Chile, and the United States of America.

\newpage
\begin{table}[!h]
\begin{center}
\begin{tabular}{|c|c|c|c|c|c|}
\hline
\hline
           &                &            &             &    &\\
$r_{int}$  &    $r_{ext}$  & $N_{BSS}$  & $N_{RGB}$   & $N_{HB}$ & $L_{samp}^{ann}$/$L_{samp}^{tot}$   \\
           &                &            &             &  & \\
\hline
\hline
    $0\arcsec$  &  $50\arcsec$    &    43    &   183   &    78 & 0.24 \\
    $50\arcsec$  & $150\arcsec$	&      46    & 	328   &    125 & 0.43 \\
   $150\arcsec$ & $300\arcsec$	&      16    & 	171   &    47  & 0.20 \\
   $300\arcsec$  & $550\arcsec$	&      3     & 	80    &    23  & 0.10 \\
  $550\arcsec$  & $1150\arcsec$	&     4     & 	24    &     7  & 0.03  \\
\hline
\hline
\end{tabular}
\end{center}
\caption{Decontaminated number counts of BSS, RGB and HB stars and fractions of sampled light.}
\label{}
\end{table}

\newpage 
 
\begin{figure}
\includegraphics[scale=0.7]{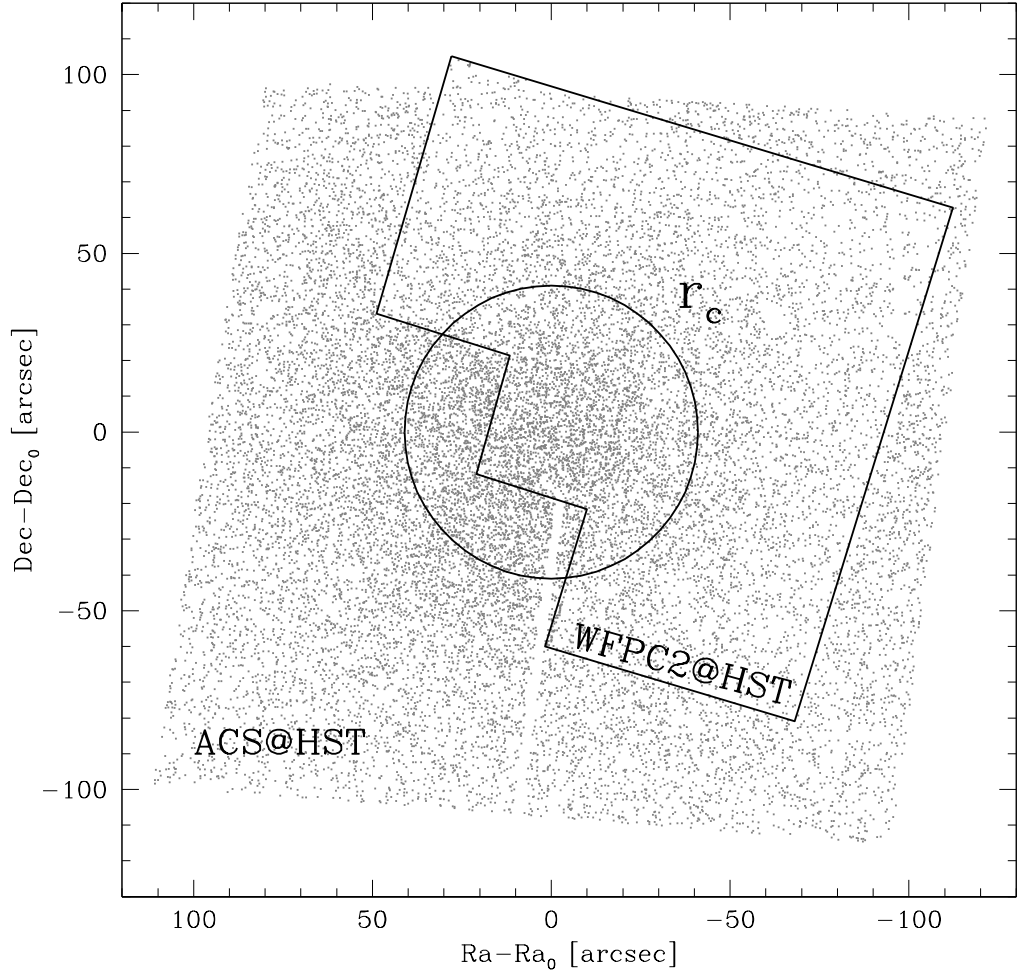}
\caption{Map of the WFPC2 and ACS data. The black circle indicates the cluster core radius $r_c$ (Section~2.3). }
\label{map}
\end{figure}

\begin{figure}
\includegraphics[scale=0.7]{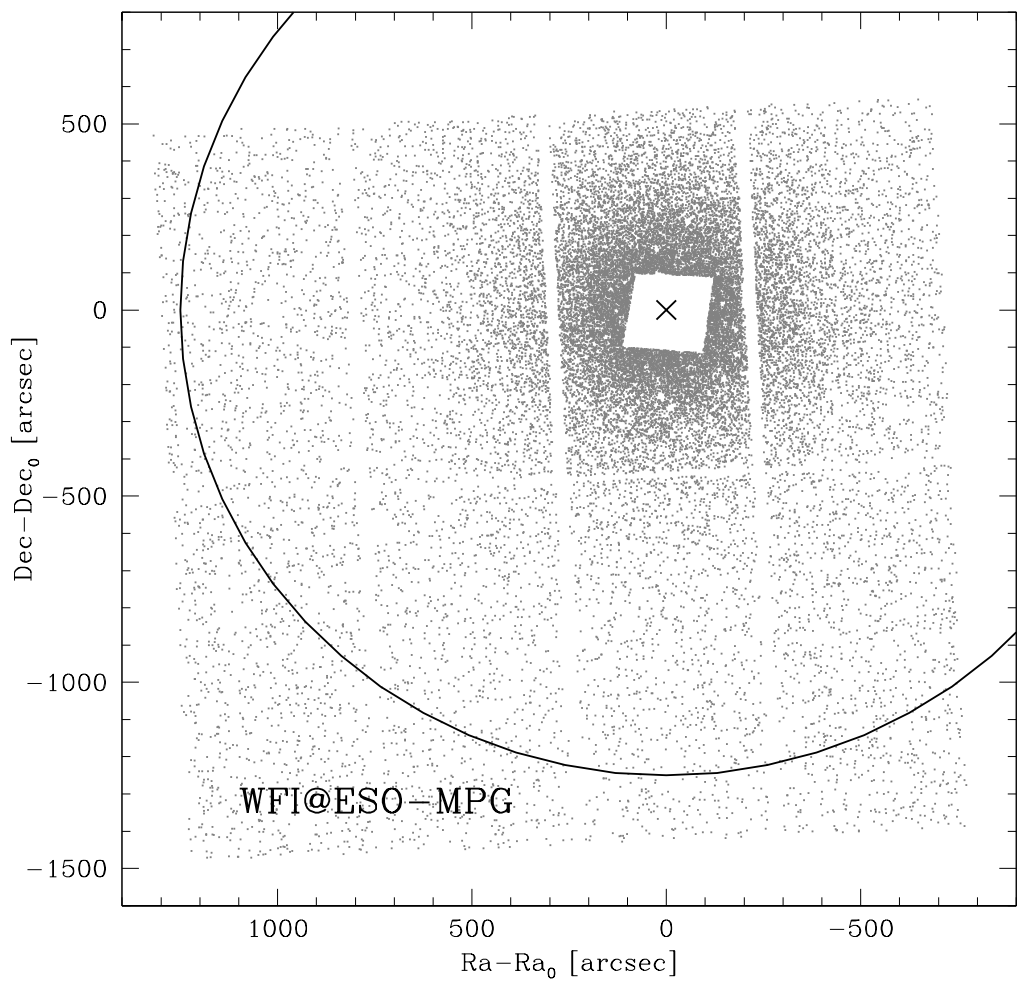}
\caption{Map of the WFI data. The blank area corresponds to the ACS and WFPC2 FOVs. The black
circle indicates the tidal radius 
$r_t$  of M10 (see Section 2.3).}
\label{map}
\end{figure}

\begin{figure}
\includegraphics[scale=0.7]{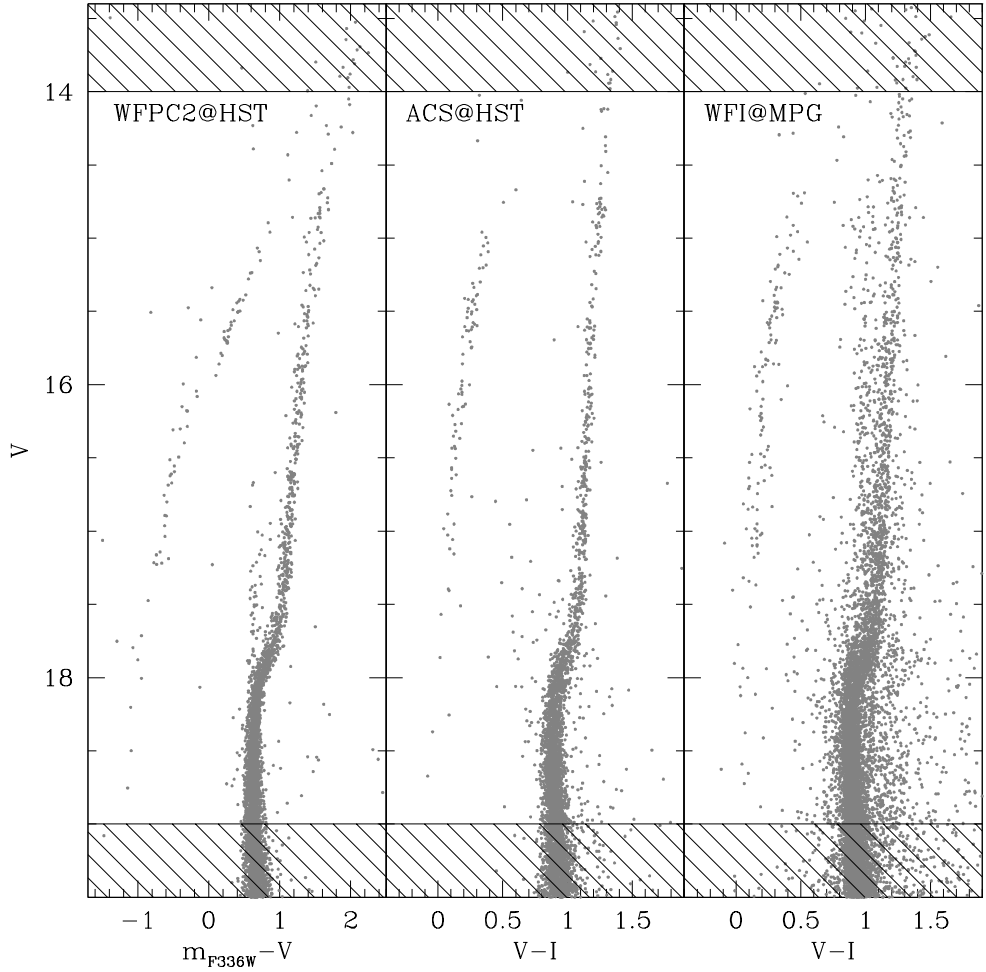}
\caption{Optical CMDs of the WFPC2, ACS and WFI samples. The shaded regions ($V<14$ and $V>19$) delimit the samples
excluded from the density profile calculation. In the rightmost panel only stars with $r<r_t$ are plotted.}
\label{map}
\end{figure}

\begin{figure}
\includegraphics[scale=0.7]{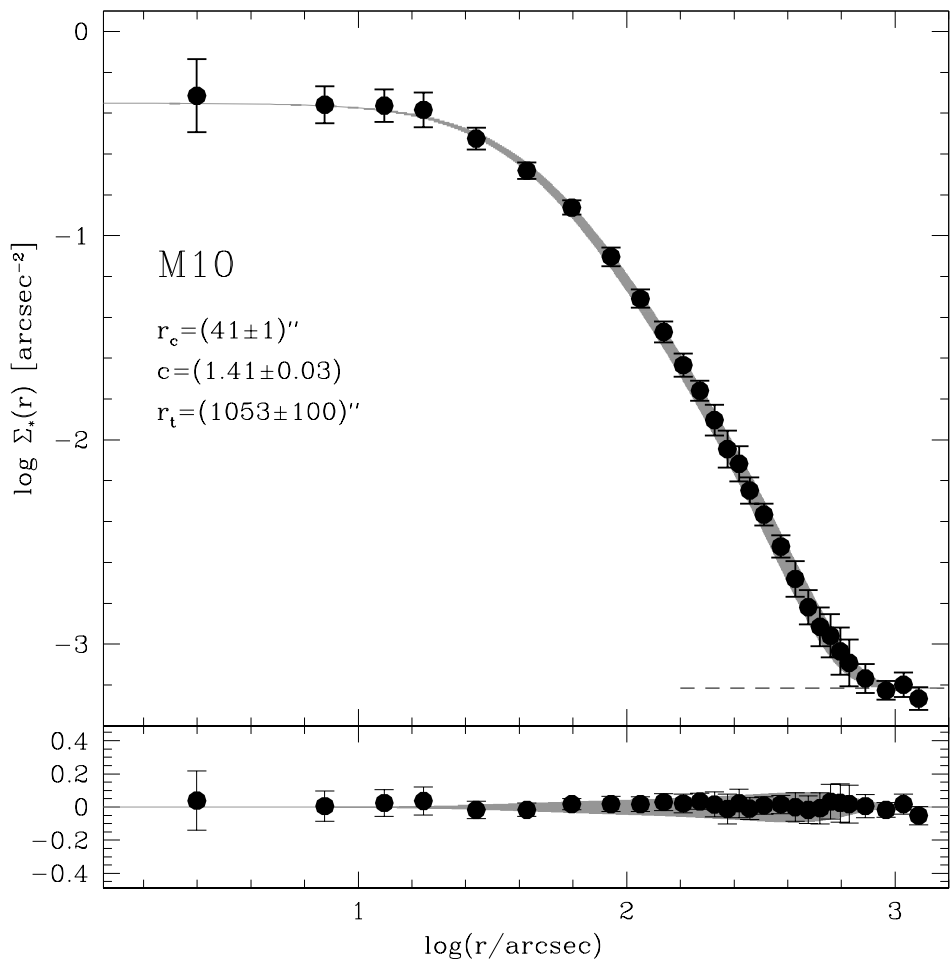}
\caption{Observed surface density profile (black dots), based on counts of all cataloged stars with $14<V<19$.
The filled grey region represents the area defined 
by the best-fit King model and the relative errors on the structural parameters. The dashed
line indicates the measured level of the background. The parameters of the King model  
(core radius and concentration) and relative uncertainties are indicated in the figure. 
The lower panel shows the residuals in each radial bin.}
\label{map}
\end{figure}

\begin{figure}
\includegraphics[scale=0.7]{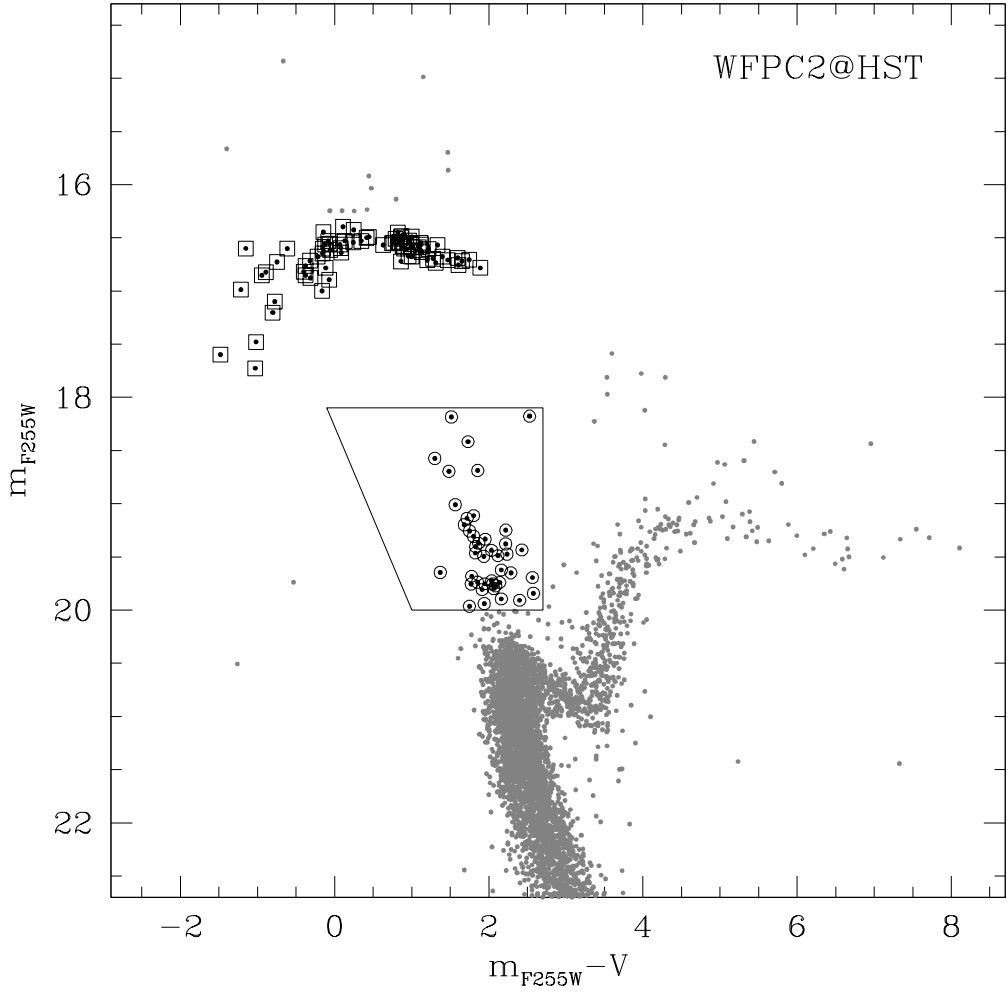}
\caption{UV CMD of the WFPC2 sample. The box shows the magnitude and color limits defined for the BSS
selection in this photometric plane. Open circles are the selected BSS, open squares are HBs.}
\label{map}
\end{figure}

\begin{figure}
\includegraphics[scale=0.7]{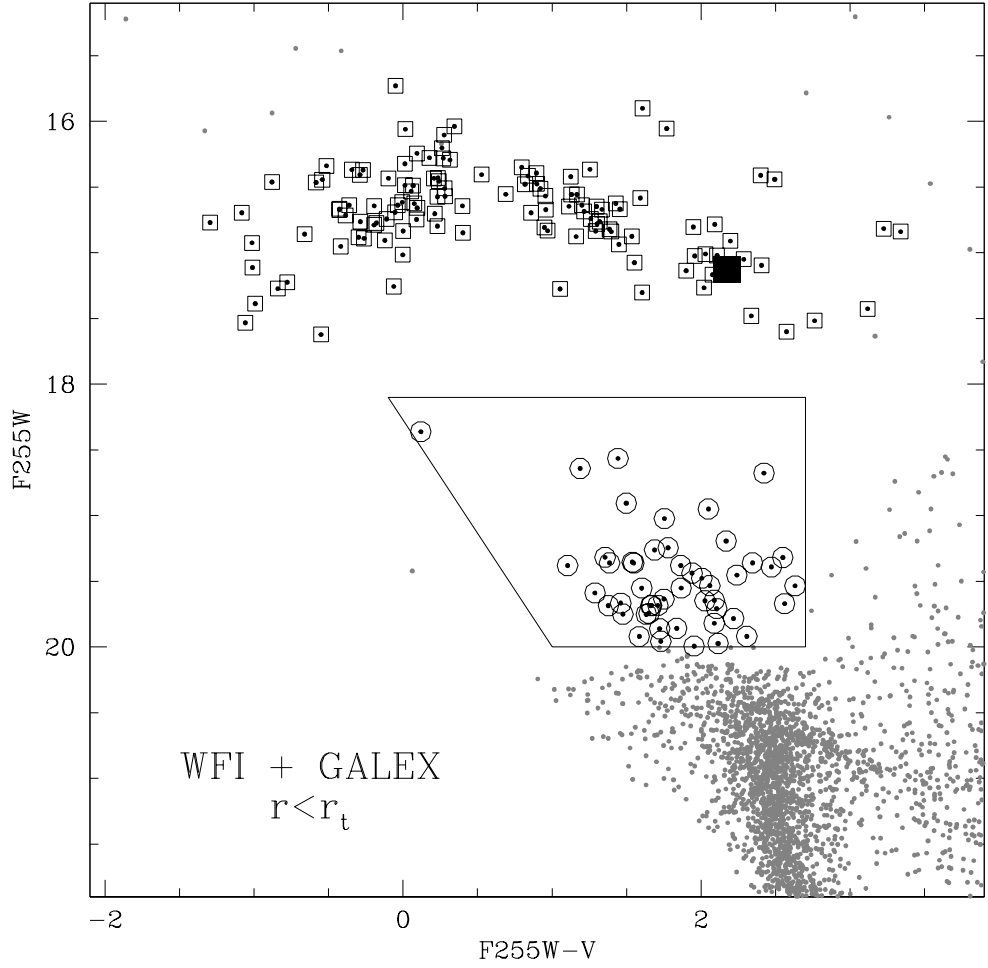}
\caption{As Figure~5, but for the WFI/GALEX sample at $r<r_t$. The large solid square marks the location of 
the Quasar [HGP92] 165429.30-040340.3. }
\label{map}
\end{figure}

\begin{figure}
\includegraphics[scale=0.7]{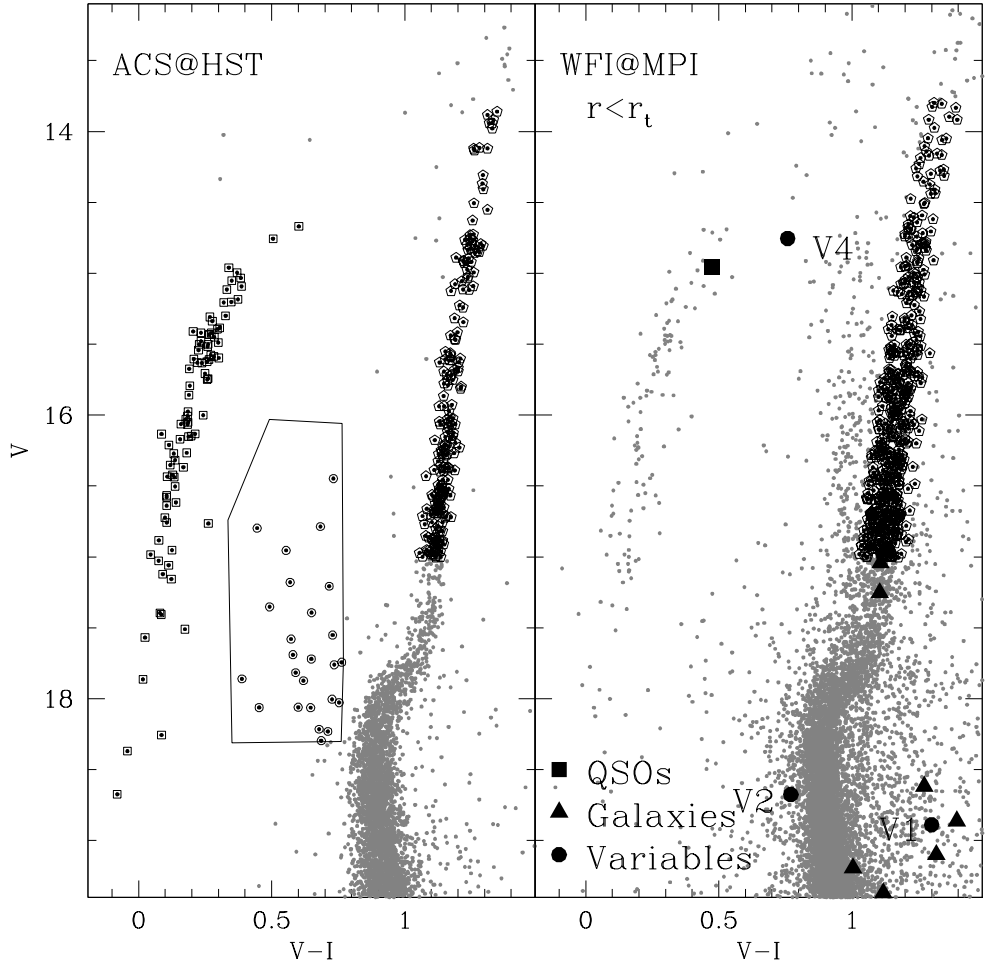}
\caption{Optical CMDs of the ACS and WFI samples, on the left and right panel respectively.
The box in the left panel defines the fiducial region for BSS selection
in the optical plane (see for details Section~3). As in Figure~5, open circles are BSS and opens squares are
HBs. Open pentagons are RGB stars. 
Quasars, galaxies and variables identified in literature (Section 3.2) are also highlighted with 
in the right panel as solid squares, triangles and circles, respectively.}
\label{map}
\end{figure}

\begin{figure}
\includegraphics[scale=0.7]{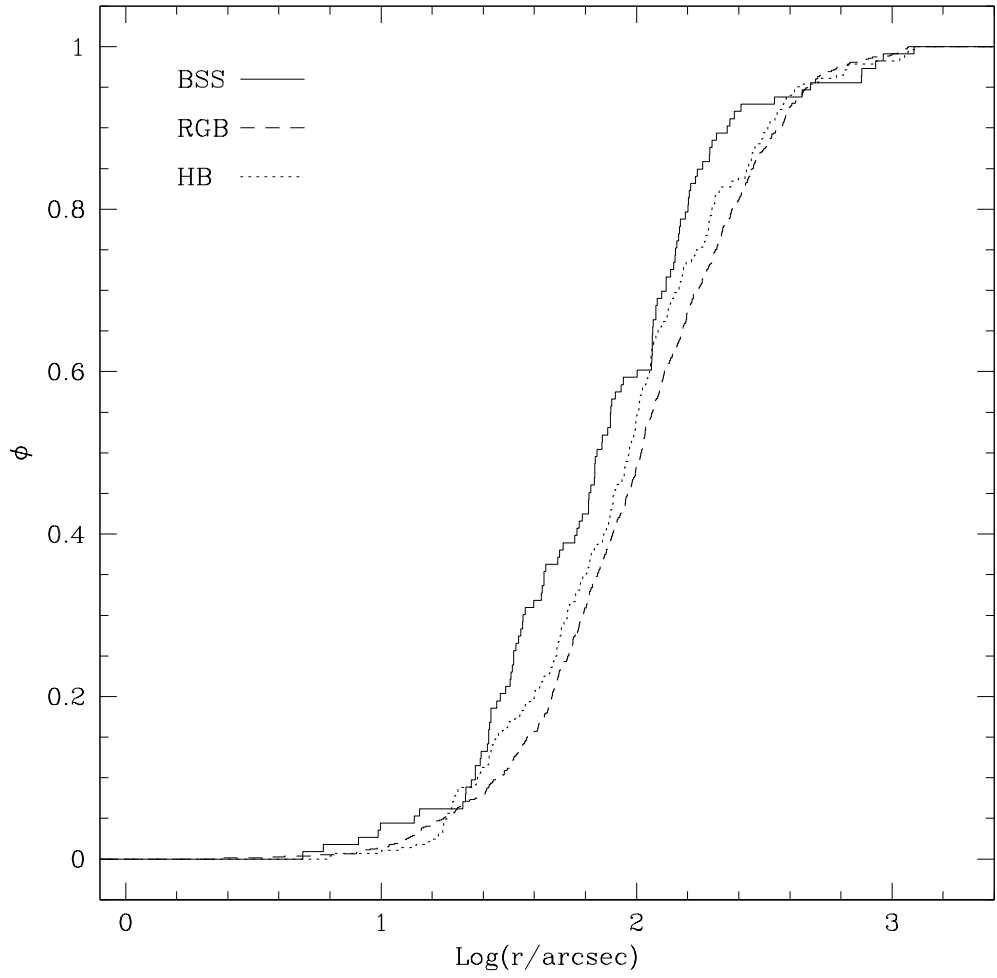}
\caption{Cumulative radial distribution of the statistically decontaminated populations.}
\label{map}
\end{figure}

\begin{figure}
\includegraphics[scale=0.7]{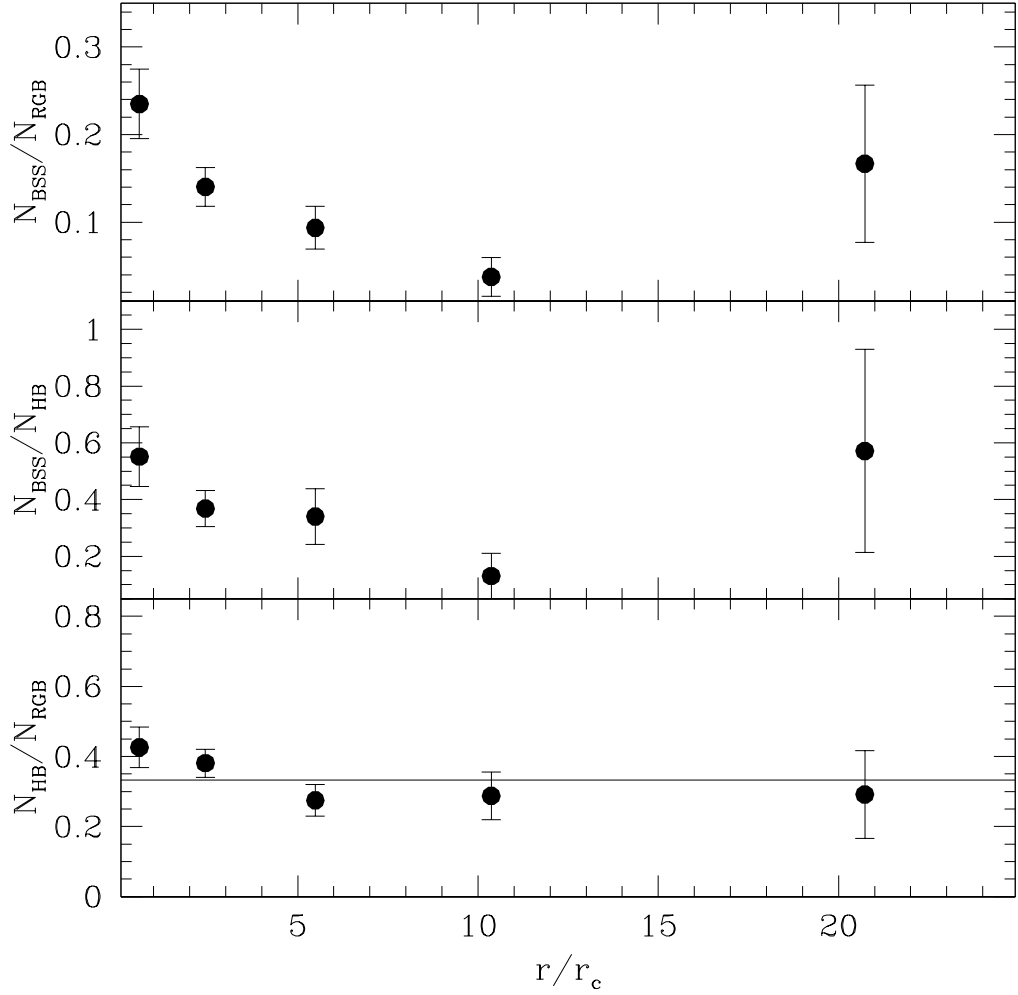}
\caption{From top to bottom, $N_{BSS}/N_{RGB}$, $N_{BSS}/N_{HB}$ and $N_{HB}/N_{RGB}$
as a function of the distance from $C_{grav}$ normalized to $r_c=41\arcsec$. The solid line in the bottom panel 
represents the average value of  $N_{HB}/N_{RGB}$ for the entire cluster extension.}
\label{map}
\end{figure}

\begin{figure}
\includegraphics[scale=0.7]{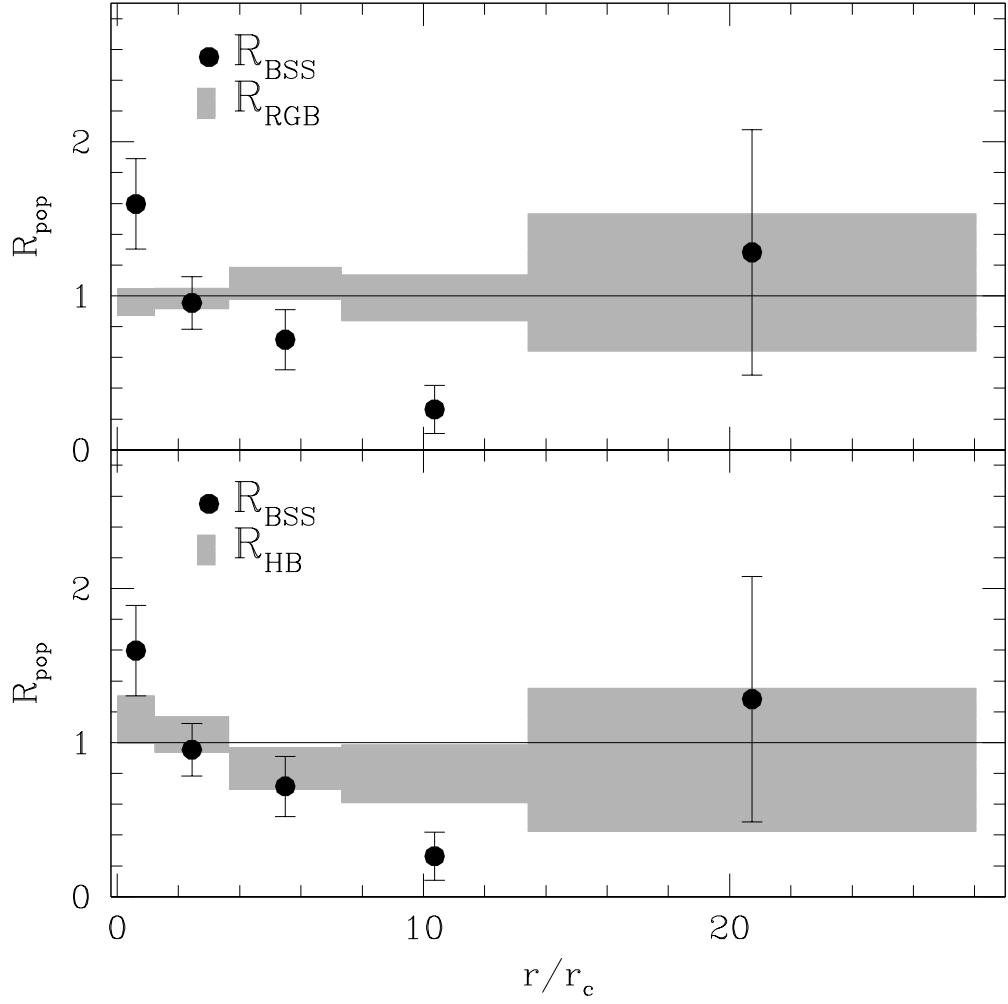}
\caption{Radial distribution of the double normalized ratios of the BSS (black dots), RGB (upper panel) 
and HB (lower panel).}
\label{map}
\end{figure}

\begin{figure}
\includegraphics[scale=0.7]{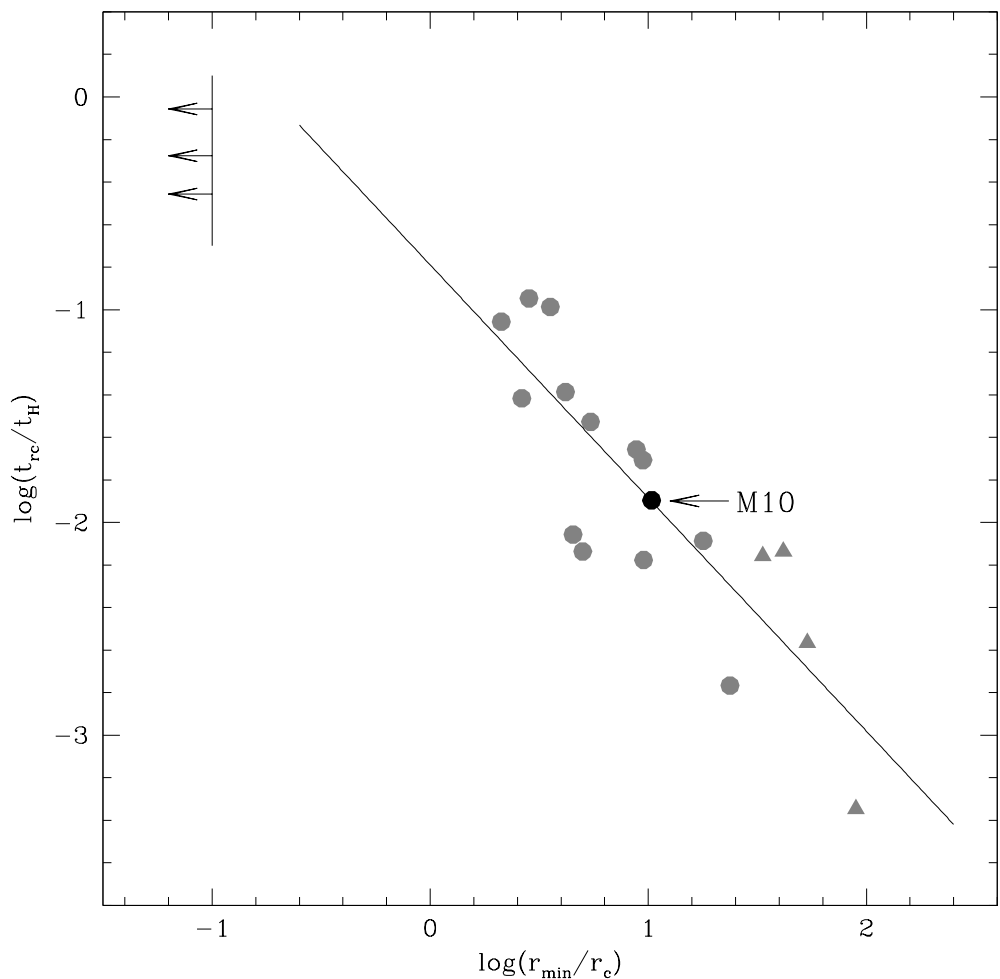}
\caption{The relaxation time at the cluster centre ($t_{rc}$) normalized to the age of the Universe ($t_H=13.7$
Gyr) as a function of $r_{min}/r_c$. This plot is the same as Figure~4 of F12. The dynamically young clusters, for
which it is not possible to define a minimum in the distribution, are shown as lower limit arrows at
$r_{min}=0.1$. Grey circles are clusters classified as {\it Family II} in F12. M10 belongs to this family and
its position has been highlighted in black. Triangles are the dynamically old clusters ({\it Family III}).}
\label{map}
\end{figure}

\end{document}